\documentclass[aps,amsmath,amsfonts,amssymb,reprint,floatfix]{revtex4-1}
\pdfoutput=1
\usepackage{graphicx}
\usepackage{graphics}
\usepackage{xcolor}
\usepackage{bm}
\begin{document}
\title{Nuclear Magnetic Resonance in Low-Symmetry Superconductors}
\author{D. C. Cavanagh}
\email{d.cavanagh@uq.edu.au}
\affiliation{School of Mathematics and Physics, The University of Queensland, Brisbane, Queensland 4072, Australia}
\author{B. J. Powell}
\affiliation{School of Mathematics and Physics, The University of Queensland, Brisbane, Queensland 4072, Australia}

\begin{abstract}
	We consider the nuclear spin-lattice relaxation rate, $1/T_1T$ in superconductors with accidental nodes. We show that a Hebel-Slichter-like peak occurs even in the absence of an isotropic component of the superconducting gap. The logarithmic divergence found in clean, non-interacting models is controlled by both disorder and electron-electron interactions. However, for reasonable parameters, neither of these effects removes the peak altogether.
\end{abstract}

\maketitle
\setlength{\unitlength}{1mm}

\section{Introduction}

Unconventional superconducting states are often defined by the breaking of an additional symmetry beyond global gauge invariance  \cite{AnnettReview,AnnettRev,SigristUeda}. Typically this means a reduction of the point group symmetry, but could also include breaking time reversal or, for triplet superconductors, spin rotation symmetries  \cite{Vollhardt&Wolfle}.

Starting from the SO(3)$\times$SO(3)$\times$U(1) symmetry of superfluid $^3$He \cite{Vollhardt&Wolfle}, the discussion of unconventional superconductivity has focused on superconductivity in high symmetry environments. In this context, unconventional superconducting states can be characterised by the irreducible representations of the point group of the crystal \cite{AnnettReview,AnnettRev,SigristUeda}.

The symmetry of the superconducting order parameter provides important clues about the mechanism of superconductivity. As such the determination of the exact form of the superconducting gap is of significant importance. In practice, such a determination is not straightforward. The Josephson interference experiments responsible for unambiguously identifying the `d$_{x^2-y^2}$-wave' symmetry of the cuprates \cite{Wollman,Tsuei} have not been possible in many materials. The interpretation of other experimental results can be  ambiguous. In particular, the limiting low temperature behaviours of many experimental probes (e.g., heat capacity, nuclear magnetic relaxation rate, or penetration depth) can, in principle, distinguish between a fully gapped state, line nodes, and point nodes. However, these results are often controversial. And, even in principle, such experiments cannot differentiate between different gap symmetries with the same class of nodes (point or line). This has led to the study of directional probes, such as thermal conductivity \cite{IzawaOrganic,IzawaSRO}. 

However, superconductivity is observed in many materials with rather low point group symmetries. Non-centrosymmetric materials are a prominent example. Here spin-orbit coupling can mix singlet and triplet superconducting states \cite{Sigrist}. Many organic superconductors, e.g., those based on the BEDT-TTF, Pd(dmit)$_2$, TMTSF, or TMTTF molecules, form monoclinic or orthorhombic crystals \cite{Ishiguro}. This means that superconducting symmetries that are distinct on the square lattice such as s-wave, d$_{xy}$, or d$_{x^2-y^2}$ often belong to the same irreducible representation \cite{BenGroupTh,BenGroupTh2}. 
Similarly, a number of transition metal oxides with orthorhombic crystal structures superconduct \cite{Dynes,Ono,YBCOPG,RotSymLSCO}. In some cuprates, chemical doping results in a distortion of the lattice, reducing the rotational symmetry to C$_2$ (i.e., orthorhombic as opposed to tetragonal). This distortion is on the order of $<10\%$ of the lattice spacing \cite{YBCOPG,RotSymLSCO}.

Emergent physics can also lower the symmetry of a material, for example, via electronic `nematicity'. Indeed, in some cuprates, even if the crystal lattice is constrained to reduce this distortion, evidence of electronic nematicity has been observed in transport properties \cite{RotSymLSCO}, while nematic phases (with reduced rotational symmetry) have been theorised, resulting from spin or charge density wave order \cite{NemCDWSDW}, and evidence of such phases, and their connection to the pseudogap phase, has been observed in some cuprate materials from magnetic torque measurements \cite{NPhys_NemThermo}. Additionally, nematic phases  arise in iron-based  superconductors \cite{Kasahara12,FernandesMillis13} (in fact, as temperature is lowered, FeSe undergoes a structural transition to an orthorhombic state well above the superconducting critical temeprature \cite{FeSeOrth}) and strong anisotropy has been observed in resistivity measurements of the heavy-fermion superconductor CeRhIn$_5$ \cite{NemInCeRhIn5}, indicating the presence of some nematic order.

Superconductivity in  materials such as the cuprate, organic,  heavy fermion and iron-based families of unconventional superconductors, is widely believed to arise from  electronic correlations. These unconventional superconductors share many similar properties, including complex phase diagrams with multiple phases and (spin singlet) superconductivity in particular proximity to some magnetically ordered phase. While the order parameter is believed to be anisotropic in the majority of these materials, disagreement remains over the exact form of the gap function  in many materials. For example, in the organic superconductors, specific heat measurements have been taken to indicate nodeless (`s-wave') superconductivity \cite{Wosnitza00,Muller02,Wosnitza03} while others indicate the presence of nodes of the gap function \cite{Nakazawa97,Taylor07,Malone10}, and similarly  penetration depth measurements were inconclusive until recently \cite{BrounkBr}. This has led both theorists \cite{Schmalian,BenGroupTh,BenRVB2,Guterding16,PowellPRL17} and experimentalists \cite{Dion,Guterding16PRL} to discuss the possibility of accidental nodes in organic superconductors. In the iron-based superconductors both `d-wave' (nodal) gap structures and nodeless `s$_\pm$-wave' structures, with band-dependent magnitudes, have been proposed for various materials \cite{ChenFeSe,Hanaguri_spm}, while in some heavy fermion superconductors a band-dependent gap symmetry has been discussed \cite{Broun2,Tanatar_HFgap} (i.e. with nodes present on some bands and isotropic gap magnitude on others).

There are important differences between accidental nodes and those required by symmetry. In the latter case, the location of the nodes is restricted to satisfy a symmetry constraint (for example, a reflection through the plane of the node line) and therefore the gap function transforms as a non-trivial representation of the point group. In the case of accidental nodes, the positioning of the nodes is unrestricted by symmetry requirements, and the gap function transforms as the trivial representation of the point group, the same representation to which an isotropic gap function belongs. This also allows the possibility of a mixed symmetry `s+d-wave' state \cite{Guterding16,Guterding16PRL}. For example, the `d$_{xy}$-wave' and `d$_{x^2-y^2}$-wave' gaps belong respectively to the B$_{2g}$ and A$_{1g}$ (trivial) representation of the D$_{2h}$ point group, which captures the orthorhombic symmetry of a square lattice with a rectangular distortion \cite{BenGroupTh,groupfoot}. As many models find d$_{x^2-y^2}$ superconductivity on the square lattice one expects that, at least for small distortions, this will also be the dominant superconducting channel for similar models  with  D$_{2h}$ point group symmetry. However, generically a real material  with this symmetry will be able to lower its energy by producing an admixture of isotropic (s-wave) superconductivity, e.g., via sub-dominant interactions. If this admixture is small it will not remove the nodes, but will move them (note that an admixture with a complex phase breaks time reversal symmetry, and so will not be considered here).

Below we consider the nuclear spin-lattice relaxation rate $1/T_1$ in superconductors with accidental nodes. We show that in clean non-interacting models there is a logarithmic divergence $1/T_1T$ as $T\rightarrow T_c$ even if there is no isotropic component of the gap, $\Delta_{\bm k}$, i.e. $\int d^3{\bm k}\Delta_{\bm k}=0$, where $T$ is the temperature and $T_c$ is the superconducting critical temperature. We show, numerically in a D$_{2h}$ symmetric model -- similar to those discussed above -- that this divergence is controlled but not removed entirely by either disorder or electron-electron interactions, giving rise to a Hebel-Slichter-like peak. However, it shows some subtle differences from the true Hebel-Slichter peak both in its microscopic origin and in that it is not controlled by gap anisotropy.

\section{Nuclear magnetic resonance and the relaxation rate $1/T_1T$}

The spins of atomic nuclei relax by exchanging energy with their environment. In the case of a metal or superconductor this means the conduction electrons.  
Thus the relaxation rate of nuclei in an electronic environment is related to the transverse dynamic susceptibility of the quasiparticles,  $\chi_{+-}\left(\bm{q},\omega\right)=\chi'_{+-}\left(\bm{q},\omega\right)+i\chi''_{+-}\left(\bm{q},\omega\right)$, via  \cite{Slichter,Tinkham}
\begin{eqnarray}
\frac{1}{T_1T}&=& \lim\limits_{\omega\rightarrow 0} \frac{2k_B}{\gamma_e^2\hbar^4}\sum\limits_{\bm{q}}\left|A_H\left(\bm{q}\right)\right|^2\frac{\chi''_{+-}\left(\bm{q},\omega\right)}{\omega},\label{T1T_Gen}
\end{eqnarray}
where $\gamma_e$ is the (electron) gyromagnetic ratio and $A_H\left(\bm{q}\right)$ is the hyperfine coupling, which we approximate by a point contact interaction [$A_H\left(\bm{q}\right)=A_H$] for simplicity below. Neglecting vertex corrections, the dynamic susceptibility can be expressed in terms of the spectral density function, $A_{\bm{k}}\left(E\right)$, \cite{Mahan,Coleman,Eddy_K}
\begin{widetext}
\begin{eqnarray}
\chi''_{+-}\left(\bm{q},\omega\right)&=&\sum\limits_{\bm{k}}\int\limits_{-\infty}^{\infty} \frac{dE_1dE_2}{4\pi^2}\left\lbrace \frac{1}{2}\left[1+\frac{\xi_{\bm{k}}\xi_{\bm{k}+\bm{q}}+\Delta_{\bm{k}}\Delta_{\bm{k}+\bm{q}}}{E_{\bm{k}}E_{\bm{k}+\bm{q}}}\right]\left[f\left( E_{2}\right)-f\left(E_{1}\right)\right]\delta\left[\omega -\left(E_{2} - E_{1}\right)\right]A_{\bm{k}}\left(E_{1}\right)A_{\bm{k}+\bm{q}}\left( E_{2}\right)\right. \nonumber\\
&&+\frac{1}{4}\left[1-\frac{\xi_{\bm{k}}\xi_{\bm{k}+\bm{q}}+\Delta_{\bm{k}}\Delta_{\bm{k}+\bm{q}}}{E_{\bm{k}}E_{\bm{k}+\bm{q}}}\right]\left[\bar{f}\left( E_{2}\right)-f\left(E_{1}\right)\right]\delta\left[\omega +\left(E_{2} + E_{1}\right)\right]A_{\bm{k}}\left(E_{1}\right)A_{\bm{k}+\bm{q}}\left( E_{2}\right)
\nonumber\\
&&\left. +\frac{1}{4}\left[1-\frac{\xi_{\bm{k}}\xi_{\bm{k}+\bm{q}}+\Delta_{\bm{k}}\Delta_{\bm{k}+\bm{q}}}{E_{\bm{k}}E_{\bm{k}+\bm{q}}}\right]\left[f\left( E_{2}\right)-\bar{f}\left(E_{1}\right)\right]\delta\left[\omega -\left(E_{2} + E_{1}\right)\right]A_{\bm{k}}\left(E_{1}\right)A_{\bm{k}+\bm{q}}\left( E_{2}\right)\right\rbrace , \label{GeneralChi1}
\end{eqnarray}
\end{widetext}
where $\xi_{\bm{k}}$ is the single particle dispersion, measured from the chemical potential, 
 $E_{\bm{k}}=\sqrt{\xi_{\bm{k}}^2+\left|\Delta_{\bm{k}}\right|^2}$, $f\left(E\right)$ is the Fermi-Dirac distribution function, and $\bar{f}\left(E\right)=1-f\left(E\right)$. The coherence factors in the above expression arise naturally from the trace over Nambu spinors.

From Eq. (\ref{GeneralChi1}) we consider three approximations: (i) The clean, BCS limit, where interactions are limited to those giving rise to superconductivity and no disorder is present. 
(ii) Uncorrelated  disorder, where electron-impurity interactions result in a broadening of the peak in the spectral function (i.e. giving rise to a finite quasiparticle lifetime). The strong disorder regime is irrelevant as strong disorder suppresses unconventional superconductivity \cite{Mineev&Samokhin}.  And (iii) disorder and electron-electron interactions, with the latter taken into account via the random phase approximation (RPA). In the following analysis we will treat the clean BCS limit analytically, then extend our results to account for disorder and electron-electron interactions  numerically.

\begin{figure}
	\includegraphics[trim = 40mm 75mm 50mm 75mm, clip, width=0.4\textwidth]{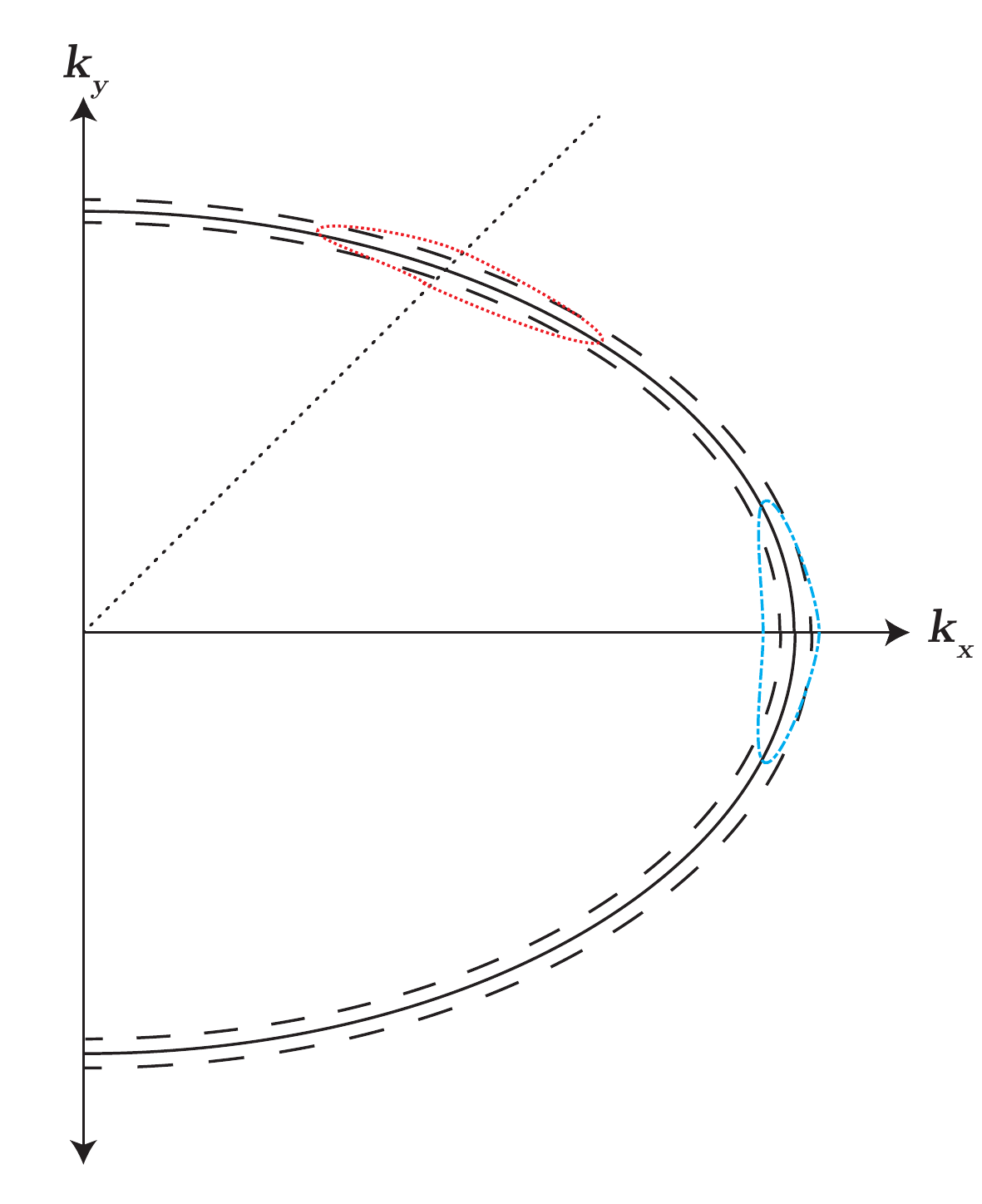}
	\caption{Contours of constant energy for an isotropic gap (black dashed lines), symmetry required nodes (blue dot dashed line) and accidental nodes (red dotted line), sketched for an elliptical Fermi surface. The black dotted line indicates the position of the accidental node, while the symmetry required nodes reside on the axes (with the contour around the $k_x$-axis highlighted).}
	\label{contourFig}
\end{figure}

\section{The clean limit}

Inserting Eq. (\ref{GeneralChi1}) for the dynamic susceptibility into the expression for the relaxation rate, Eq. (\ref{T1T_Gen}), we have
\begin{eqnarray}
\frac{1}{T_1T} &\propto &\frac{1}{4\pi^2}\int\limits_{-\infty}^{\infty} dE\left[-\frac{df}{dE}\right]
\sum_{n=1}^3
\left[\mathcal{K}_n\left(E\right)\right]^2  ,\label{GeneralKernel}
\end{eqnarray}
where 
\begin{subequations}
\begin{eqnarray}
\mathcal{K}_1\left(E\right)&=&\sum_{\bm{k}}A_{\bm{k}}\left(E\right),\\
\mathcal{K}_2\left(E\right)&=&\sum_{\bm{k}}\frac{\xi_{\bm{k}}}{E_{\bm{k}}}A_{\bm{k}}\left(E\right),\\
\mathcal{K}_3\left(E\right)&=&\sum_{\bm{k}}\frac{\Delta_{\bm{k}}}{E_{\bm{k}}}A_{\bm{k}}\left(E\right).
\end{eqnarray}
\label{eq:Kdef}
\end{subequations}
Each term in Eq. (\ref{GeneralKernel}) represents the average of a function over a contour of approximately constant energy $E_{\bm{k}}$, due to the peak in $A_{\bm k}(E)$ at $E\simeq E_{\bm k}$. These averages are then integrated over energy, with the integral restricted by the derivative of the Fermi function to a range of order $k_BT$. For an s-wave superconductor, this contour will wrap around the entire Fermi surface, while for a gap with nodes, the contour will form closed surfaces around the nodes (for energies smaller than the maximum gap). Examples of these contours are shown in Fig. \ref{contourFig}.

\begin{figure}
			\includegraphics[trim = 50mm 105mm 45mm 45mm, clip, width=0.4\textwidth]{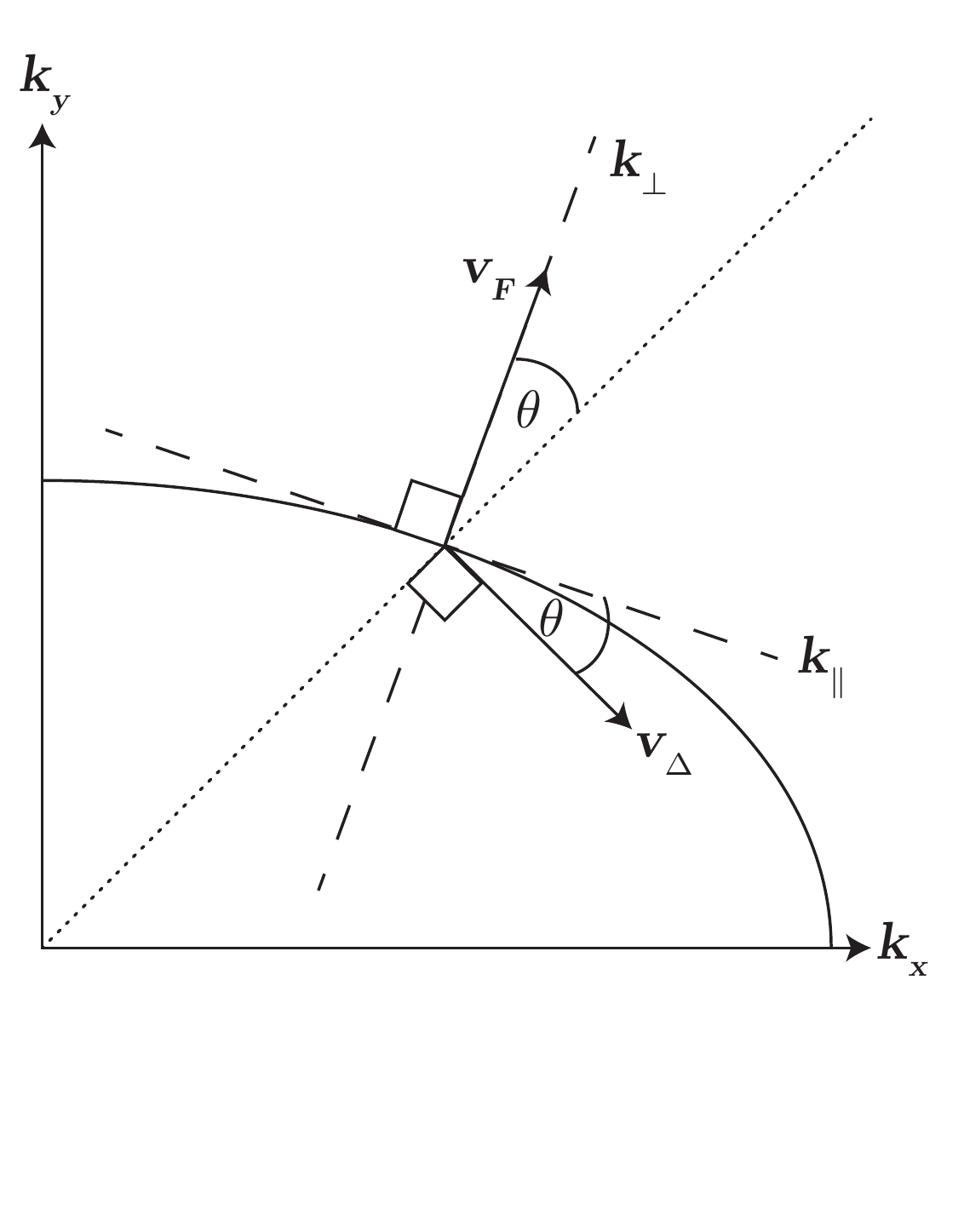}
		\caption{The geometry of a system with accidental nodes. Arrows denote the gradients of the normal dispersion ($v_F$) and the gap function ($v_\Delta$), in the local co-ordinate system defined at the node. The solid line denotes the Fermi surface and the dotted line the node location, while $k_\parallel$ and $k_\perp$ respectively denote the coordinates parallel and perpendicular to the Fermi surface.}
		\label{angleFig}
\end{figure}

\subsection{Anisotropic gap with accidental nodes}\label{accidental}
In the clean limit, the spectral functions are given by Dirac delta functions,
\begin{eqnarray}
A_{\bm{k}}\left(E\right)&=&\pi\delta\left(E_{\bm{k}}-E\right), \label{Spectral}
\end{eqnarray}
 each of which can be decomposed  \cite{Arfken,G&V} into delta functions acting on $k_\perp$, the component of $\bm{k}$ perpendicular to the Fermi surface,
\begin{eqnarray}
\delta\left(E_{\bm{k}}-E\right)&=&\sum\limits_{\bm k_\perp\left(E\right)}\left|\frac{\partial E_{\bm{k}}}{\partial \bm k_\perp}\right|^{-1}\delta\left[\bm k_{\perp}-\bm k_{\perp}\left(E\right)\right].
\end{eqnarray} 
These delta functions then constrain $\bm k_\perp$ to the value corresponding  to the constant energy surface, $\bm k_\perp\left(E\right)$. The gradient of the gap function is then, in terms of 
$\bm{k}_\perp$ and a local $\left(D-1\right)$-dimensional manifold parallel to the Fermi surface, $\{\bm{k}_i\}$, given by
\begin{eqnarray}
\bm{v}_{\Delta}\equiv\boldsymbol{\nabla}_{\bm{k}}\Delta_{\bm{k}}&=&\sum\limits_{i=1}^{D-1}\hat{\bm{k}}_{i}\left(v_{\Delta}\right)_i +\hat{\bm{k}}_{\perp}v_{\perp},
\end{eqnarray}
where $\left(v_{\Delta}\right)_i=\bm{v}_{\Delta}\cdot\bm k_i$ is the projection of $\boldsymbol{\nabla}_{\bm{k}}\Delta_{\bm{k}}$ onto $\bm{k}_i$, the $i$th dimension on the manifold parallel to the Fermi surface, $v_{\perp}=\bm{v}_{\Delta}\cdot\bm k_\perp$ is the projection of $\boldsymbol{\nabla}_{\bm{k}}\Delta_{\bm{k}}$ onto $\bm{k}_\perp$, $\bm k_i=\hat{\bm k}_ik_i$, and  $\bm k_\perp=\hat{\bm k}_\perp k_\perp$. In general, the $(v_{\Delta})_i$ and $v_{\perp}$  are functions of the position on the manifold. In the simplest, two dimensional, case (with $\bm{k}_i=\bm{k}_\parallel$), this gives 
\begin{eqnarray}
\bm{v}_{\Delta}\equiv\boldsymbol{\nabla}_{\bm{k}}\Delta_{\bm{k}}&=&\hat{\bm{k}}_{\parallel} v_{\Delta} \cos\theta 
+\hat{\bm{k}}_{\perp}v_{\Delta} \sin\theta ,
\end{eqnarray}
where $\theta $ is the angle between the nodal line and the normal to the Fermi surface (see Fig. \ref{angleFig}).

In general, the gap function will be independent of the coordinate parallel to the node line. Importantly, in the accidental case this is not required to be normal to the Fermi surface, which results in a nonvanishing average of the gap over the Fermi surface. The Hebel-Slichter peak present in $1/T_1T$ in an isotropic superconductor is  a probe of this average gap \cite{AnnettRev}. The angle $\theta$  parametrises the existence of such a nonvanishing average gap, and can be defined via the overlap between the gradients of the gap and dispersion, as the dispersion varies solely in the direction normal to the Fermi surface,
\begin{eqnarray}
\sin\theta =\frac{\bm{v}_F\cdot \bm{v}_{\Delta}}{\left|\bm{v}_F\right|\left|\bm{v}_{\Delta}\right|},\label{sinth_def}
\end{eqnarray} 
where $\bm{v}_{F}=\boldsymbol{\nabla}_{\bm{k}}\xi_{\bm{k}}$ is the Fermi velocity and $\theta=\phi+\pi/2$, where $\phi$ is the angle between $\bm{v}_{\Delta}$ and $\bm{v}_F$. Near the node the energy is then $E_{\bm{k}} \sim \sqrt{\left[v_Fk_{\perp}\right]^2+v_\Delta^2\left[k_\parallel \cos\theta + k_{\perp}\sin\theta \right]^2}$. 

The delta function in Eq. (\ref{Spectral}) {then constrains the component perpendicular to the Fermi surface, allowing the simplification, 
\begin{widetext}
\begin{eqnarray}
\delta\left[E_{\bm{k}} - E\right]&=&\sum\limits_{k_\perp\left(E\right)}\left|\frac{E}{v_F\sqrt{E^2-\left|\Delta_{\bm{k}}\right|^2}+\Delta_{\bm{k}}v_{\Delta} \sin\theta }\right|\delta\left[k_\perp-k_\perp\left(E\right)\right].\label{myDeltaExp}
\end{eqnarray}
Previous analyses \cite{Tinkham,Samokhin} have primarily focused on the symmetry required case. A symmetry required node must reside on an axis of symmetry, to which the Fermi surface must be perpendicular; thus $\theta=0$.  In fact, we find that the vanishing of this angle is responsible for the lack of analytical divergences encountered in the symmetry required case, see Section \ref{sect:req}.

As a first approximation, we perform a binomial expansion in $\Delta_{\bm{k}}v_{\Delta}\sin\theta/v_F\sqrt{E^2-\left|\Delta_{\bm{k}}\right|^2}$ in the denominators in Eqs. (\ref{eq:Kdef}).  Such an approximation is valid for $T\sim T_c$, where $\Delta_{\bm{k}}\rightarrow 0$  provided $v_\Delta/v_F$ is not too large; i.e., away from van Hove singularities, where $v_F$ vanishes. Additionally, for sufficiently small $\sin\theta $, such an expansion will be reasonable at all temperatures given the same caveat.
 Performing the expansion gives
\begin{subequations}
\begin{eqnarray}
\mathcal{K}_1\left(E\right) &=& \int_{E}d\bm{k}_\parallel \frac{E}{v_F\sqrt{E^2-\Delta_{\bm{k}}^2}+\Delta_{\bm{k}}v_{\Delta} \sin\theta }\approx \int_{E}d\bm{k}_\parallel \left[\frac{E}{v_F\sqrt{E^2-\Delta_{\bm{k}}^2}}-\frac{E\Delta_{\bm{k}}v_{\Delta} \sin\theta}{v_F^2\left(E^2-\Delta_{\bm{k}}^2\right)}\right]\label{Terms1}\\
\mathcal{K}_2\left(E\right) &=& \int_{E}d\bm{k}_\parallel \frac{\sqrt{E^2-\Delta_{\bm{k}}^2}}{v_F\sqrt{E^2-\Delta_{\bm{k}}^2}+\Delta_{\bm{k}}v_{\Delta} \sin\theta }\approx \int_{E}d\bm{k}_\parallel \left[\frac{1}{v_F}-\frac{\Delta_{\bm{k}}v_{\Delta} \sin\theta }{v_F^2\sqrt{E^2-\Delta_{\bm{k}}^2}}\right]\label{Terms2}\\
\mathcal{K}_3\left(E\right) &=& \int_{E}d\bm{k}_\parallel \frac{\Delta_{\bm{k}}}{v_F\sqrt{E^2-\Delta_{\bm{k}}^2}+\Delta_{\bm{k}}v_{\Delta} \sin\theta } \approx \int_{E}d\bm{k}_\parallel\left[ \frac{\Delta_{\bm{k}}}{v_F\sqrt{E^2-\Delta_{\bm{k}}^2}}-\frac{\Delta_{\bm{k}}^2v_{\Delta} \sin\theta }{v_F^2\left(E^2-\Delta_{\bm{k}}^2\right)}\right],\label{Terms3}
\end{eqnarray}
\label{allTerms}
\end{subequations}
\end{widetext}
where $\int_{E}d\bm{k}_\parallel$ denotes the integral over the $\left(D-1\right)$-dimensional surface in momentum space at energy $E$. 

In the terms depending on the gap, we approximate the gap function by a Taylor series in the momentum components parallel to the Fermi surface, near the node, in $D$-dimensions this is given by
\begin{eqnarray}
\Delta_{\bm{k}}&=\sum\limits_{i=1}^{D-1}\left(v_{\Delta}\right)_i  \left(k^i_\parallel-k^{i\left(0\right)}_\parallel\right)+\mathcal{O}\left(k^i_\parallel-k^{i\left(0\right)}_\parallel\right)^2, \label{NearNodegen}
\end{eqnarray} 
where $\bm{k}_\parallel^{\left(0\right)}=\hat{\bm{k}}_\parallel^{\left(0\right)} k_\parallel^{\left(0\right)}$ denotes the position of the node. In the case of $D=2$ this gives
\begin{eqnarray}
\Delta_{\bm{k}}&=&v_{\Delta}\cos\theta  \left(k_\parallel-k_\parallel^{\left(0\right)}\right)+\mathcal{O}\left(k_\parallel-k_\parallel^{\left(0\right)}\right)^2. \label{NearNode2D}
\end{eqnarray} 

Under this approximation for the gap, we arrive at 
\begin{subequations}
\begin{eqnarray}
\mathcal{K}_1\left(E\right)&=&  \lim\limits_{\delta\rightarrow 0}\frac{E}{2}\left\langle\frac{\text{sgn}\left(\Delta\right)}{v_F\cos\theta}\left[\frac{\pi }{v_{\Delta}} +\frac{\sin \theta }{v_F}\ln \left(\delta\right)\right]\right\rangle_{E}\label{K1acc}\hspace{0.7cm}\\
\mathcal{K}_2\left(E\right)&=& \left\langle v_F^{-1}\right\rangle_{E}\label{K2acc}\\
\mathcal{K}_3\left(E\right)&=& \lim\limits_{\delta\rightarrow 0}\frac{E}{2}\left\langle\frac{\tan \theta }{v_F^2}\right\rangle_{E}\ln \left(\delta\right),
\label{K3acc}
\end{eqnarray}
\end{subequations}
where $\left\langle \ldots \right\rangle_{E}$ denotes the average over the  contour(s) of energy $E$.
Note that $\mathcal{K}_1\left(E\right)$ depends on the difference between the averages taken on the segments of the energy contour with positive and negative superconducting gap, while $\mathcal{K}_2$ and $\mathcal{K}_3$ depend only on averages over the entire energy contour. The logarithmically divergent contributions arise due to the vanishing denominators in the expansion terms of Eqs. (\ref{Terms1}) and (\ref{Terms3}), while the terms with square roots in the denominator give a convergent contribution.
 
In the general case of accidental node placement, the velocity magnitudes $v_{\Delta}$ and $v_{F}$ may  vary freely across the energy surface, but as an demonstrative example, in the simplest case, $v_F$ and $v_{\Delta}$ are constant near the node, giving 
\begin{eqnarray}
\mathcal{K}_1\left(E\right)&=&0\label{Ccdef}\\
\mathcal{K}_3\left(E\right)&=&\frac{E}{2}\lim\limits_{\delta\rightarrow 0}\left\langle\frac{\tan \theta }{v_F^2}\right\rangle_{E}\ln \left(\delta\right)\label{Bcdef}.
\end{eqnarray}

Both the linear correction to Eq. (\ref{Terms2}) and the zero order contribution to Eq. (\ref{Terms3}) vanish as the gap function is odd with respect to the position of the node, but higher order corrections will diverge, similar to the divergence encountered in Eq. (\ref{K3acc}). In a more realistic model, the velocities $v_F$ and $v_\Delta$, as well as the angle $\theta$ will depend on $\bm{k}$, and so $\mathcal{K}_1\left(E\right)$ may also be nonvanishing. In general, the node in an anisotropic gap function is not required to be near a portion of the gap function where a linear expansion in $k_\parallel$ is valid, in particular the presence of an isotropic component of the gap will shift the node position. Including higher order terms in either the Taylor series for the gap or the expansions of the $\mathcal{K}_i$ is, however, insufficient to remove these divergences.

\subsection{Anisotropic gap with symmetry required nodes}\label{sect:req}

If $\Delta_{\bm{k}}$ transforms as a non-trivial representation of the point group, the nodes are required by symmetry. This implies that $\theta $ vanishes, as the gap function near the node is independent of the direction perpendicular to the Fermi surface. Further, as the node in this case is required to reside on a symmetry axis for the material, $\mathcal{K}_3$ must vanish, given that an equal length of the contour is on either side of the node where the gap changes sign. In this case, $\mathcal{K}_2$ [Eq. (\ref{Terms2})] again gives a non-divergent contribution with the form of the density of states, as the surface integral over the energy contour; and Eq. (\ref{Terms1}) reduces to $\mathcal{K}_1\left(E\right)=\pi E/\left\langle v_{\Delta} v_F\right\rangle$. In this way, we recover the well known result \cite{Coleman,Ketterson&Song} that no Hebel-Slichter peak is observed.

\subsection{Isotropic gap}
In a purely isotropic gap superconductor, the gap function is independent of momentum, so $\bm{v}_\Delta=0$ and $\Delta_{\bm{k}}=\Delta_0$, and the momentum sums in the relaxation rate give the constant energy surface integral $S\left(E\right)=\sum_{\bm{k}_\parallel ,k_\perp\left(E\right)}\frac{1}{v_F\left(\bm{k}\right)}\delta\left[k_\perp -k_\perp\left(E\right)\right]$, 
\begin{eqnarray}
\mathcal{K}_1\left(E\right) &=& \int_{E}d\bm{k}_\parallel \frac{E}{v_F\sqrt{E^2-\Delta_{0}^2}}=\frac{E}{\sqrt{E^2-\Delta_{0}^2}}S\left(E\right)\\
\mathcal{K}_2\left(E\right) &=& \int_{E}d\bm{k}_\parallel \frac{\sqrt{E^2-\Delta_{0}^2}}{v_F\sqrt{E^2-\Delta_{0}^2} }=S\left(E\right)\\
\mathcal{K}_3\left(E\right) &=& \int_{E}d\bm{k}_\parallel \frac{\Delta_{0}}{v_F\sqrt{E^2-\Delta_{0}^2} }=\frac{\Delta_0}{\sqrt{E^2-\Delta_{0}^2}}S\left(E\right).\hspace*{0.6cm}
\end{eqnarray}
Thus, $\mathcal{K}_2$ is once more  non-divergent. 
The energy integrals over $\mathcal{K}_1^2$ and $\mathcal{K}_3^2$ result in a logarithmic divergence in the energy domain when $E\sim T \rightarrow\Delta$, giving rise to the Hebel-Slichter peak. 
The divergence is in general controlled by one or more of gap anisotropy, the presence of impurities or strong coupling effects  \cite{Tinkham, Ketterson&Song, Samokhin}. 

This is in marked contrast to the accidental node case, where the peak results from a divergence in the momentum integral, independently of the energy value. The peak observed in the isotropic case arises due to a divergence at a particular energy. 

\section{Disorder and Electron-Electron Interactions}

The classical s-wave Hebel-Slicter peak is controlled by disorder, electron-electron interactions, and gap anisotropy. The latter is necessarily present in the case of accidental nodes. Therefore, it is important to understand the affects of the former two effects in the current context.   

If we evaluate the energy integrals in Eq. (\ref{GeneralChi1}), first using the delta function to constrain $E_2$ before using the strongly peaked nature of the spectral function to evaluate $E_1\approx E_{\bm{k}}$, we have
\begin{widetext}
\begin{eqnarray}
\chi''_{+-}\left(\bm{q},\omega\right)&=&\sum\limits_{\bm{k}}\left\lbrace \frac{1}{2}\left[1+\frac{\xi_{\bm{k}}\xi_{\bm{k}+\bm{q}}+\Delta_{\bm{k}}\Delta_{\bm{k}+\bm{q}}}{E_{\bm{k}}E_{\bm{k}+\bm{q}}}\right]\left[f\left(E_{\bm{k}}+\omega\right)-f\left(E_{\bm{k}}\right)\right]A_{\bm{k}+\bm{q}}\left( E_{\bm{k}}+\omega\right)\right. \nonumber\\
&&+\frac{1}{4}\left[1-\frac{\xi_{\bm{k}}\xi_{\bm{k}+\bm{q}}+\Delta_{\bm{k}}\Delta_{\bm{k}+\bm{q}}}{E_{\bm{k}}E_{\bm{k}+\bm{q}}}\right]\left[f\left(E_{\bm{k}}+\omega\right)-f\left(E_{\bm{k}}\right)\right]A_{\bm{k}+\bm{q}}\left( -E_{\bm{k}}-\omega\right)
\nonumber\\
&&\left. +\frac{1}{4}\left[1-\frac{\xi_{\bm{k}}\xi_{\bm{k}+\bm{q}}+\Delta_{\bm{k}}\Delta_{\bm{k}+\bm{q}}}{E_{\bm{k}}E_{\bm{k}+\bm{q}}}\right]\left[f\left( \omega -E_{\bm{k}}\right)-\bar{f}\left(E_{\bm{k}}\right)\right]A_{\bm{k}+\bm{q}}\left( \omega - E_{\bm{k}}\right)\right\rbrace , \label{NumChi2}
\end{eqnarray}
\end{widetext}
where the remaining spectral function is evaluated numerically, by introducing a finite Lorentzian broadening, of width $\eta$. This  is equivalent to introducing a finite lifetime to the familiar BCS susceptibility \cite{Coleman,Ketterson&Song,Scalapino92}

To evaluate the relaxation rate, Eq. (\ref{T1T_Gen}), each of the nested momentum integrals (over $\bm{k}$ and $\bm{q}$) are performed numerically on a discrete grid, with a small finite frequency, which is then reduced until further variations no longer affect the result, allowing the limit $\omega\rightarrow 0$ to be approximated numerically. 

\subsection{Orthorhombic  model}\label{analytical}

In order to investigate this behaviour numerically, we consider a simple tight-binding model on an orthorhombic lattice, with nearest neighbour couplings ($t_x$ and $t_y$) allowed to vary independently. The dispersion relation is thus
\begin{eqnarray}
\varepsilon_{\bm{k}} &=& t_x\cos k_x+t_y\cos k_y, \label{dispersion}
\end{eqnarray}
where we have set $a_x=a_y=1$ (where $a_x$ and $a_y$ are the lattice constants in the $x$ and $y$ directions, respectively). 

We consider two different symmetry states, a d$_{x^2-y^2}$ gap, with nodes located at $k_y=\pm k_x$, given by
\begin{eqnarray}
\Delta^{\left(x^2-y^2\right)}_{\bm{k}} &=& \frac{\Delta_0}{2}\left(\cos k_x-\cos k_y \right),\label{dgap}
\end{eqnarray}
and a d$_{xy}$ gap, with nodes located on the axes ($k_x=0$ and $k_y=0$)
\begin{eqnarray}
\Delta^{\left(xy\right)}_{\bm{k}} &=& \Delta_0\sin k_x \sin k_y.\label{xygap}
\end{eqnarray}
In both cases, the maximum magnitude of the gap is  $|\Delta_0|$. The presence or absence of a divergent peak in the $1/T_1T$ relaxation rate is dependent on the angle between the quasiparticle group velocity and the `gap velocity' at the position of the node on the Fermi surface. This angle gives an approximate measure of the value of the superconducting gap averaged over the Fermi surface, which in turn controls the presence of a Hebel-Slichter divergence. In a material with symmetry required nodes, the angle vanishes, $\theta^{(xy)}=0$ for this model, as does the average of the gap, resulting in an absence of the Hebel-Slichter peak. 

Near the $k_x=k_y$ node of the d$_{x^2-y^2}$ symmetry gap, we find

\begin{eqnarray}
\sin\theta^{(x^2-y^2)}&=& \frac{\left(t_x - t_y\right)}{\sqrt{2}\sqrt{t_x^2+t_y^2}}.
\label{theta_x2y2}
\end{eqnarray}

The above calculations estimate the parameters relevant to the divergence for specific `d-wave' gap symmetries, as these are the focus in many families of unconventional superconductor, especially the cuprate and organic superconductors. If the nodes of the gap are accidental, by definition there is no preferred node placement on the Fermi surface and the above case is  fine tuned. To explore the possibilities of other node locations, we include a finite isotropic component into the gap function. This results in a shift of the node position on the Fermi surface, while retaining the symmetry properties of the fully anisotropic gap. Additionally, such an isotropic component will alter the magnitude of the average gap on the Fermi surface, unless such effects are negated by the shifted node position. As an example, we consider a d$_{x^2-y^2}$-wave gap with an isotropic component parametrised by a real coefficient, $\alpha$, given by
\begin{eqnarray}
\Delta_{\bm{k}}\left(\alpha\right) &=& \Delta_0\left[\alpha +\left(1-\left|\alpha\right|\right)\frac{\cos k_x-\cos k_y}{2}\right],\nonumber\\\label{s+dwave}
\end{eqnarray}
where $\alpha = \pm1$ corresponds to the conventional isotropic gap, $\alpha = 0$ corresponds to the situations described in the previous section, and the absolute value of $\alpha$ is taken in the prefactor to the second term so that the magnitude of the maximum gap remains constant. We do not consider complex $\alpha$ as this would break time reversal symmetry and is hence detectable by other methods \cite{SigristUeda,BenGroupTh}. Hence,
\begin{widetext}
\begin{eqnarray}
\sin\left[\theta(\alpha , k_y)\right]&=& \frac{\left(t_x - t_y\right)\sin^2 k_y-\frac{4 t_x\alpha}{\left|\alpha\right| -1}\left[\frac{\alpha}{\left|\alpha\right| -1}-\cos k_y\right]}
{\sqrt{2}\sqrt{\left[\left(t^2_x + t_y^2\right)\sin^2 k_y
		-\frac{4 t_x \alpha}{\left|\alpha\right|-1}\left(\frac{\alpha}{\left|\alpha\right|-1}-\cos k_y\right)\right]\left[\sin^2 k_y -\frac{2\alpha}{\left|\alpha\right|-1}\left(\frac{\alpha}{\left|\alpha\right|-1}-\cos k_y \right)\right]}}.\label{theta_s+dwave}
\end{eqnarray}
\end{widetext}
Notably, this expression is now explicitly dependent on $k_y$, and therefore the shape and size of the Fermi surface, unlike in the $\alpha=0$ case considered previously. Additionally, it can be seen that, while the isotropic component may enhance the peak in the accidental node case, it can also potentially reduce the peak, depending on the relative magnitudes of the anisotropy $t_y/t_x$, the isotropic component $\alpha$ and the size and position of the Fermi surface, via $k_y$.

Eq. (\ref{theta_s+dwave}) also indicates that, even in the limit of vanishing anisotropy in the hopping parameters ($t_x\rightarrow t_y$), there arises a divergence in the relaxation rate due to the second term in the numerator for finite $\alpha$. This is entirely expected, as the gap has a non-zero average value over the Fermi surface for non-zero $\alpha$. In terms of the angle $\theta$, this can be interpreted as the isotropic component altering the nodal structure of the gap in the Brillouin zone, deforming the surface upon which nodes exist.

We take the temperature dependence of the gap to be given by given by the strong coupling BCS form:
\begin{eqnarray}
\Delta_0\left(T\right)=\frac{\Delta_0}{2}\tanh\left(3\sqrt{\frac{T_c}{T}-1}\right)
\end{eqnarray} with $\Delta_0/2=2.5k_BT_c=0.25t$, typical of a number of unconventional superconductors \cite{BrounkBr,DHS}. 

\subsection{Robustness of the Hebel-Slichter-like peak}

\begin{figure}
	\includegraphics[trim =   5mm 70mm 5mm 80mm, clip, width=0.45\textwidth]{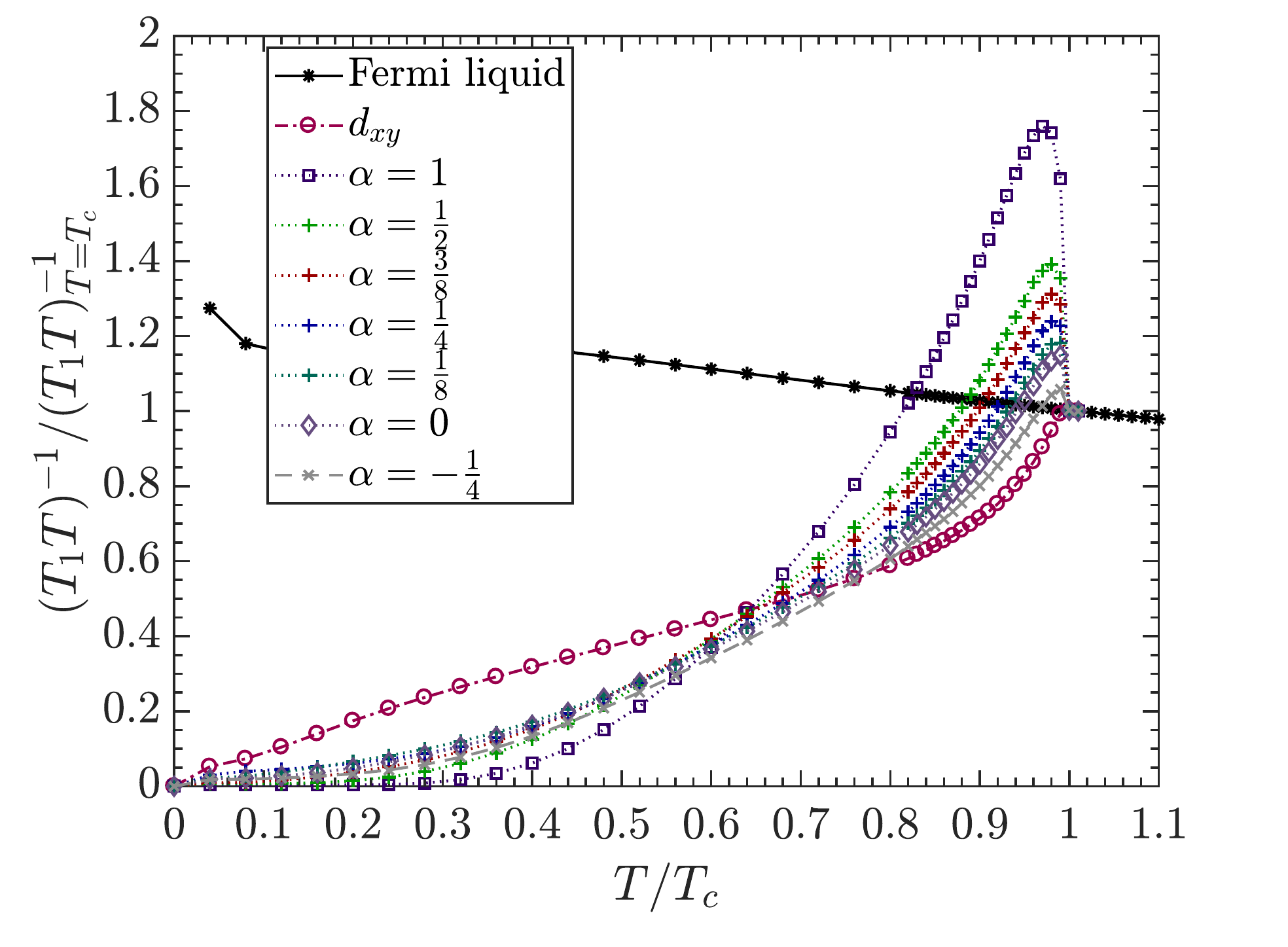}%
	\\
	\includegraphics[trim =   5mm 70mm 5mm 80mm, clip, width=0.45\textwidth]{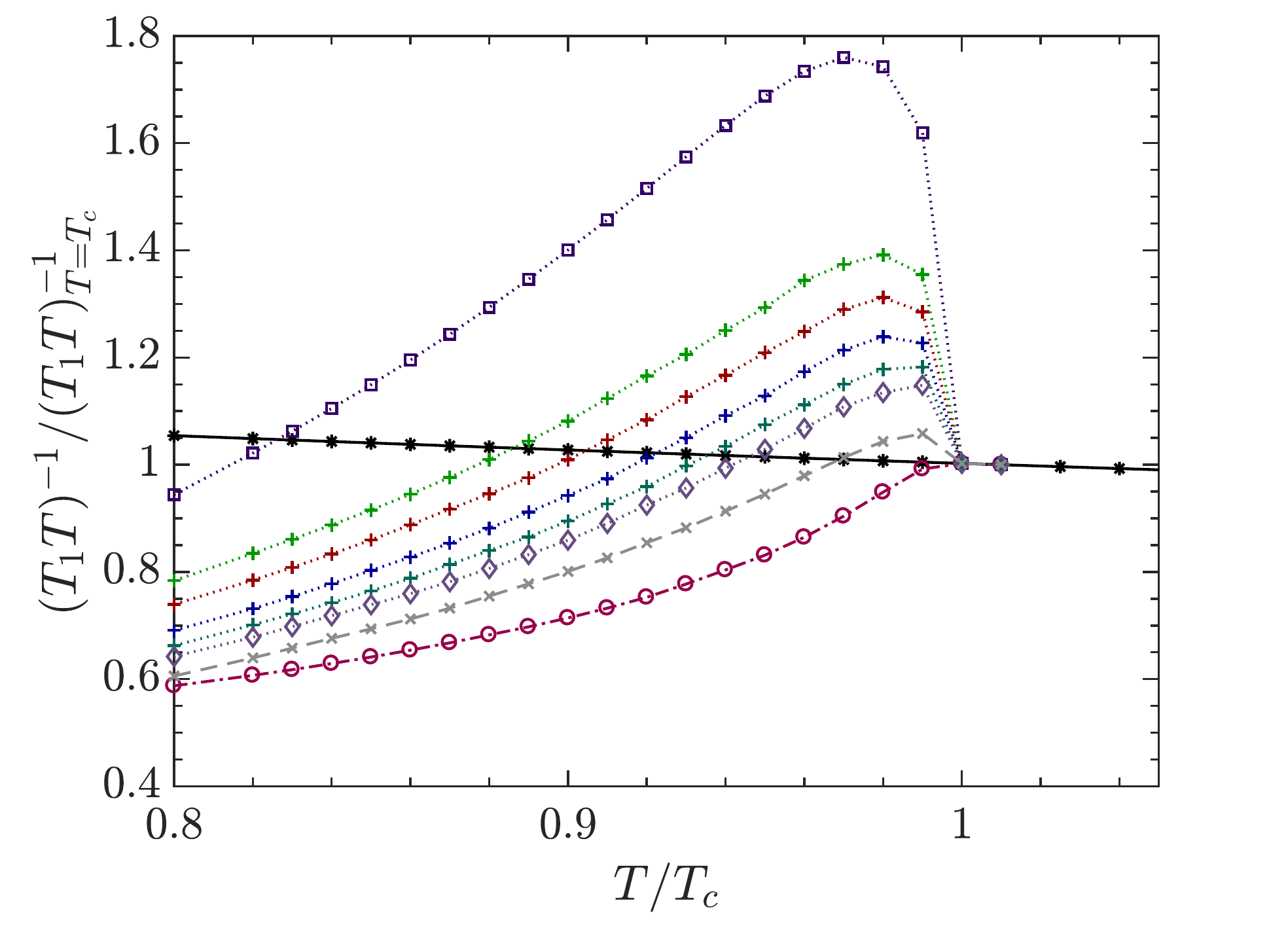}%
	\caption{Peak structure in the presence of disorder. The divergence observed in the clean limit, Eq. (\ref{K1acc}-\ref{K3acc}), is controlled by the introduction of disorder, but a clear peak remains even in the limit of large disorder. Top: Orthorhombic model, Eq. (\ref{dispersion}), with $t_y= 0.4t_x$. The relaxation rates in both the isotropic s-wave ($\alpha=1$) and symmetry required ($d_{xy}$) gap cases match conventional expectations with a Hebel-Slichter peak and its absence, respectively. In the accidental node case, we see a peak present at $\alpha=0$, which grows smoothly to the s-wave magnitude with increasing isotropic component. Furthermore, the variation of the peak is also smooth for  $\alpha <0$. Interestingly this  decreases the peak magnitude, as the angle is decreased in this case, cf. Eq. (\ref{theta_s+dwave}). Bottom: The same data, close to $T_c$, highlighting the peak structure. For these plots, frequency $\omega=5\times 10^{-3}t$, Lorentzian broadening $\eta=10^{-3}t$ (corresponding to a residual resistivity of order $\sim 10~\Omega$\,cm for $a_x,a_y\sim 3$\,\r{A}, relevant to cuprates and other transition metal oxides, up to $\sim 100~\Omega$\,cm for organic materials, with $a_x,a_y\sim 10$\,\r{A}, well above measured values in irradiated crystals \cite{Analytis}), number of grid points $N=300^4$ ($300$ per dimension in the $\bm{q}$ and $\bm{k}$ integrals) and $\left\langle n \right\rangle=0.5$ (quarter filling).}
	\label{orthopt4}
\end{figure}

In Fig. \ref{orthopt4}, we show the results of the numerical calculations for the above model with $t_y=0.4t_x$, for various gap symmetries at quarter filling. 
In the symmetry required ($\theta=0$) case, $1/T_1T$ decreases immediately below T$_c$, never increasing above the Fermi liquid value, as expected. 
For the isotropic gap ($\alpha=1$) and gaps with accidental nodes ($-1/2\leq\alpha\leq1/2$) the logarithmic divergence found in the pure case is controlled by the introduction of disorder for all gaps studied.  
Nevertheless, we find clear Hebel-Slichter-like peaks for all of the gaps with accidental nodes studied, indicating that the essential physics of this effect survives even quite strong disorder. 

It is interesting to note that the size of the peak varies smoothly with $\alpha$, cf. Eq. (\ref{s+dwave}). In particular the case $\alpha=0$, where there is no isotropic component in the gap, is not special. Indeed the peak is smaller for $\alpha<0$ than it is for $\alpha=0$. This is a straightforward consequence of the anisotropy of the Fermi surface. For $\alpha=-1/4$ the average gap over Fermi surface is less than the average for $\alpha=0$. As $\alpha$ is further decreased this average must vanish and then increase  again with the peak for $\alpha=-1$ being identical to that for $\alpha=1$.

\begin{figure}
	\includegraphics[trim =   10mm 75mm 10mm 60mm, clip, width=0.45\textwidth]{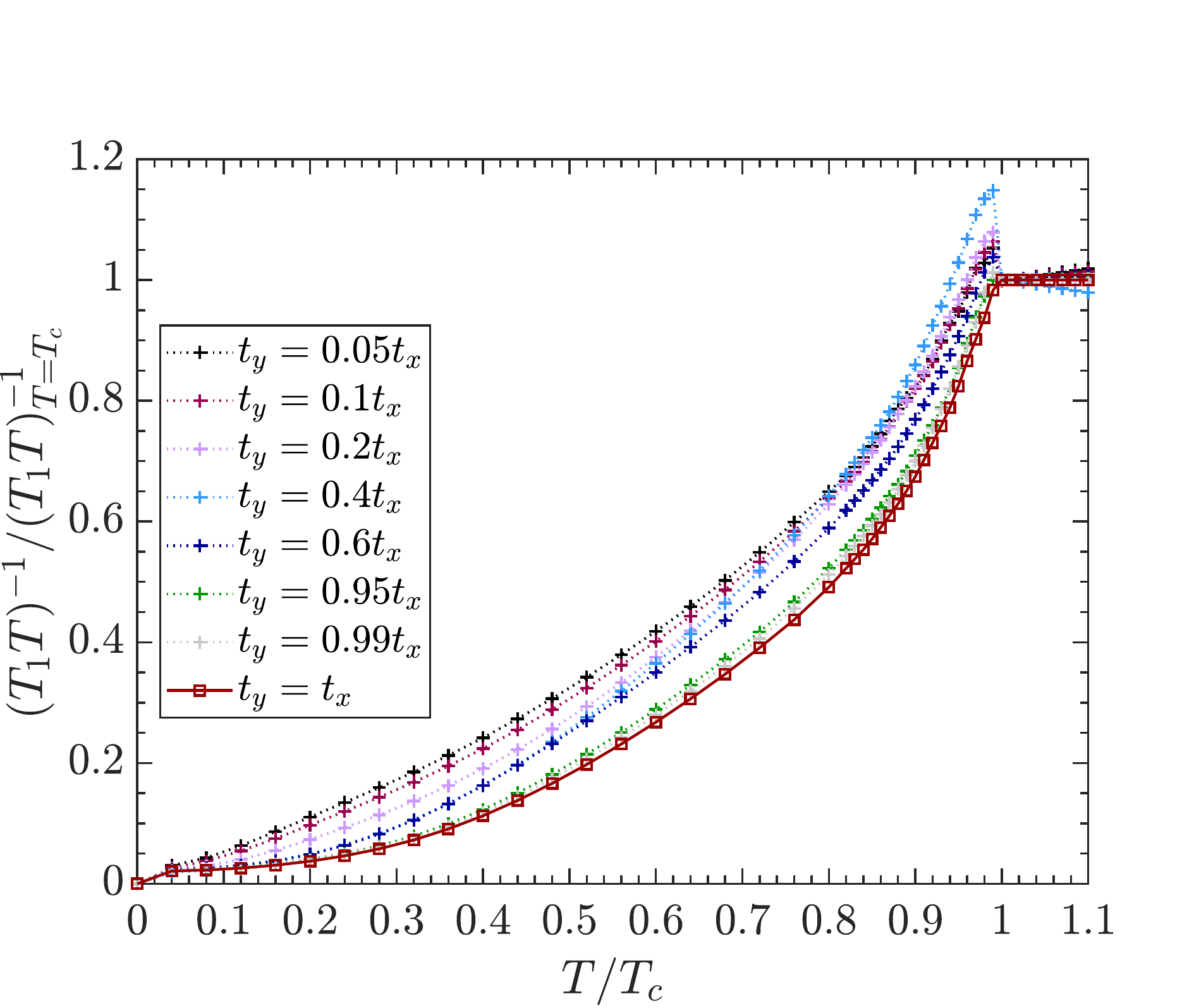}%
	\caption{Effect of the band structure anisotropy on the relaxation rate $1/T_1T$. Here we plot the calculated $1/T_1T$ for the orthorhombic model, Eq. (\ref{dispersion}), for various values of $t_y/t_x$ for the case of accidental nodes with no isotropic component ($\alpha=0$). The magnitude of the peak initially grows with increasing anisotropy, reaching a maximum value for $t_y=0.4t_x$, before decreasing again. The initial growth arises from the increase in $\theta^{(x^2-y^2)}$, cf. Eq. (\ref{theta_x2y2}). The suppression of the Hebel-Slichter-like peak for $t_y<0.4t_x$ is caused by the proximity to a van Hove singularity when the Fermi surface crosses the Brillouin zone boundary. 
		Notably, this behaviour is only seen close to $T_c$, where contours with energy $\sim k_BT$ wrap around a significant portion of the Fermi surface. At sufficiently low temperatures $1/T_1T$ also increases monotonically with increasing anisotropy (decreasing $t_y/t_x$). Note that in the normal state $1/T_1T$  depends on the hopping anisotropy, as visible from the spread of the data above $T_c$.
		 Parameters: $\omega=5\times 10^{-3}t$, $\eta=10^{-3}t$, $N=300^4$ and $\left\langle n \right\rangle=0.5$. }
	\label{orthotcomp}
\end{figure}

To better understand the dependence of the peak magnitude on the Fermi surface anisotropy  we show the $\alpha=0$ accidental node case for varying hopping anisotropy in Fig. \ref{orthotcomp}. These numerical results should be compared to the analytical prediction  that $\sin\theta \propto t_x-t_y$, Eq. (\ref{theta_x2y2}). 

At low temperatures increasing the anisotropy always increases $1/T_1T$, consistent with the changes in $\theta$. For weak anisotropies the peak grows, consistent with this prediction. However, a maximum is reached at $t_y=0.4t_x$, further increasing the anisotropy (decreasing $t_y$) decreases the peak immediately below $T_c$. This behaviour is not explained by the variation of $\theta$. 

The supression of the Hebel-Slichter-like peak for $t_y<0.4t_x$ is due to the presence of a van Hove singularity in the density of states which approaches the the Fermi energy at quarter filling as $t_y$ is reduced. 
Close to $T_c$ the gap is small, $k_BT\gtrsim \Delta_0\left(T\right)$, and contours with energy $\sim\mu\pm k_BT$ wrap around a large segment of the Fermi surface. As a result, such contours include the region of the Fermi surface where the van Hove singularity is relevant, enhancing the spectral weight (density of states) in this region. This, in turn, affects the  average of the gap within $\sim k_BT$ of the Fermi surface. In the example considered here, the superconducting gap in the vicinity of the van Hove singularity is of the minority sign of the gap, and thus the van Hove singularity reduces the average gap value over the Fermi surface. In the orthorhombic model, the van Hove singularity arises as the Fermi surface crosses the Brillouin zone boundary ($k_y=\pm\pi$), enhancing the contribution for $\Delta_{\bm{k}} <0$. As the accidental nodes are on the $k_x=\pm k_y$ diagonals, the average of the gap within $\sim k_BT$ of the Fermi surface  $\left\langle \Delta_{\bm{k}}\right\rangle_{\mu\pm k_BT}
>0$ is then reduced by the contribution due to the van Hove singularity.

Thus, for temperatures close to $T_c$, the enhancement of the spectral density at the van Hove point becomes significant, while it is less relevant at lower temperatures where the contours of energy $\sim k_BT$ are further from the van Hove point. 
 Such behaviour is not apparent from variation of $\theta$ (see Sec. \ref{accidental}), as the binomial expansion in the derivation of Eqs. (\ref{allTerms}) fails due to the divergence of $1/v_F$ near the van Hove point. The importance of such singularities are, however, apparent from Eqs. (\ref{eq:Kdef}). 

In the low temperature regime, where the gap is maximal, the relevant contours are restricted to be near the nodes, well away from the van Hove singularity, and thus the relaxation rate increases smoothly as a function of decreasing $t_y$. In the regime of smaller anisotropy ($t_y\ge0.4t_x$), the effects of the van Hove singularity are not strong enough to overwhelm the effects due to the variation of $\theta$, and the peak size increases smoothly with decreasing $t_y$.

For all levels of  anisotropy ($t_y<t_x$), we find the qualitative features observed in the $t_y=0.4t_x$ case largely unchanged, though at very low anisotropy ($t_y\gtrsim 0.95t_x$) the $\alpha = 0$ peak is strongly suppressed and not clearly resolved in the numerics.

\begin{figure}
	\includegraphics[trim =   10mm 75mm 10mm 60mm, clip, width=0.45\textwidth]{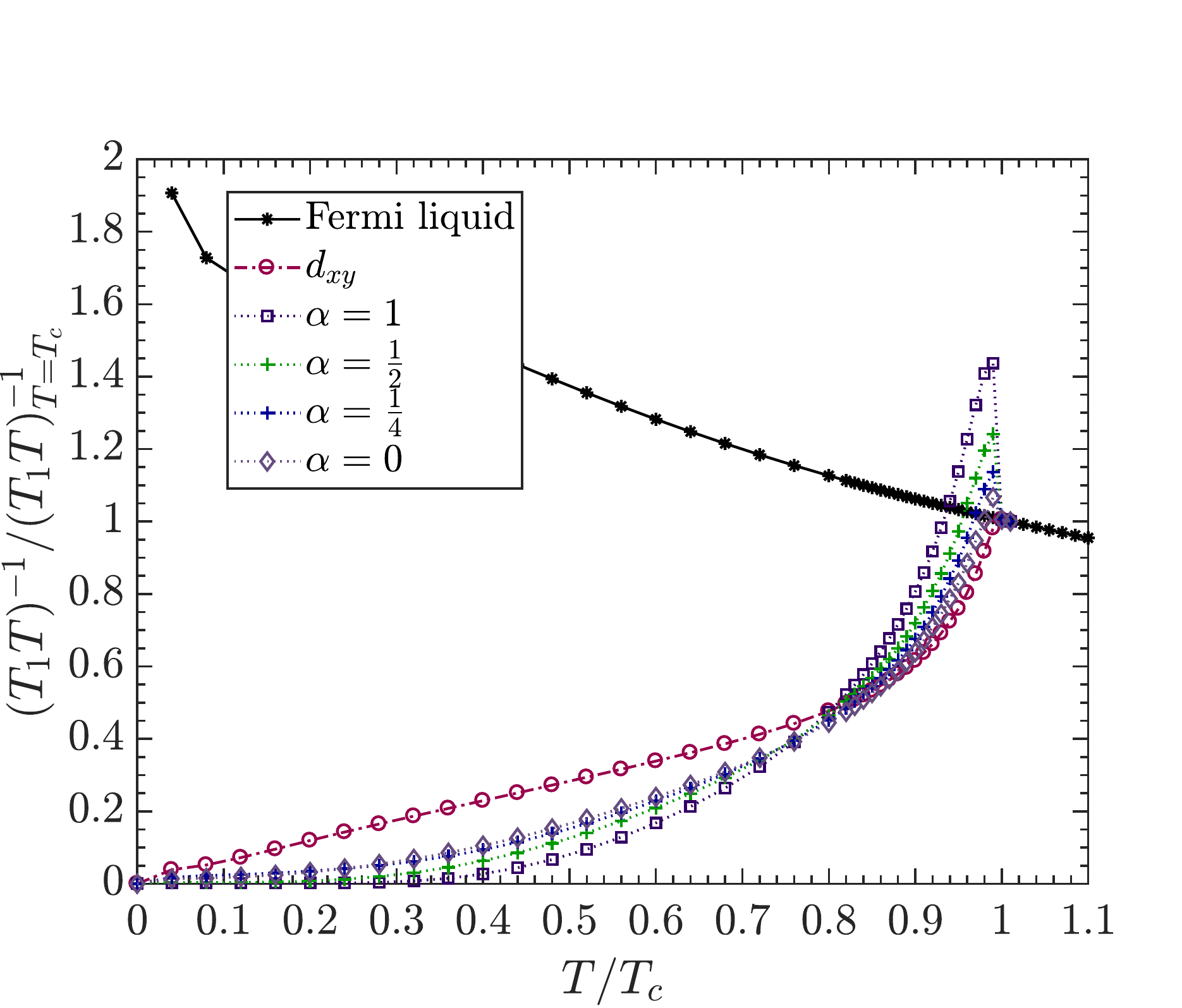}%
	\caption{Robustness of the accidental node peak to electron-electron interactions. Orthorhombic model, Eq. (\ref{dispersion}), with $t_y= 0.4t_x$, and $U=2t$ [Eq. (\ref{RPA})]. It is apparent here that the inclusion of electron-electron interactions via the RPA susceptibility does not alter the qualitative features of the previous figures. A clear Hebel-Slichter-like peak is still apparent for all values of $\alpha$ in the accidental node case, though the width of said peaks is reduced, even in the s-wave case ($\alpha=1$). The Fermi liquid relaxation rate also acquires a much stronger temperature dependence.	Parameters:  $\omega=5\times 10^{-3}t$, $\eta=10^{-3}t$, $N=300^4$ and $\left\langle n \right\rangle=0.5$. }
	\label{orthoRPA}
\end{figure}

Finally, to investigate the effects of including electron-electron interactions, we present results for the random phase approximation. The RPA for the magnetic susceptibility is the sum over ladder diagrams \cite{Doniach&Sondheimer}, therefore this treatment includes the vertex corrections that we have neglected above. Explicitly, we replace the magnetic susceptibility by
\begin{eqnarray}
\chi_{RPA}\left(\bm{q},\omega\right)&=&\frac{\chi_{+-}\left(\bm{q},\omega\right)}{1-U\chi_{+-}\left(\bm{q},\omega\right)},\label{RPA}
\end{eqnarray}
where $\chi_{+-}\left(\bm{q},\omega\right)$ is the magnetic susceptibility  (in either the superconducting or normal state, as appropriate) in the absence of electron-electron interactions. For simplicity we limit our treatment to a Hubbard-like model with a contact interaction, $U$. 

As shown in Fig. \ref{orthoRPA}, the qualitative features of the relaxation rate survive the inclusion of vertex corrections via the RPA susceptibility. Nevertheless it is important to note that the RPA treatment predicts that electron-electron interactions tend to suppress the Hebel-Slichter-like peak.

Beyond vertex corrections electron-electron interactions lead to a temperature dependence for the quasiparticle lifetime. Including such effects, for example via the phenomenological form described in \cite{Jacko}, does not lead to significant changes in $1/T_1T$.

\section{Conclusions}

We have shown that there is a logarithmic divergence in $1/T_1T$ in superconductors with accidental nodes as $T\rightarrow T_c$ from below. The microscopic origin of this divergence is distinct from that of the Hebel-Slichter peak familiar from s-wave superconductors. One signature of this is that the anisotropy in the gap, necessary for accidental nodes, does not control the divergence, as it does in the Hebel-Slichter case. We have confirmed that both impurities and electron-electron interactions can control the divergence, but for reasonable values these effects no not completely suppress the effect.

Thus, we predict a Hebel-Slichter-like peak should be observed in superconductors with accidental nodes. This provides an important test for theories of superconductivity in low symmetry materials that predict the presence of accidental nodes.

\section*{Acknowledgements}

This work was supported by the Australian Research Council (FT13010016 and DP160100060).


\begin{thebibliography}{99}
\bibitem{groupfoot} For monoclinic materials, such as the organic superconductor $\kappa$-(BEDT-TTF)$_2$Cu(NCS)$_2$, the relevant point group is the C$_{2h}$ group, in which case the `d$_{x^2-y^2}$-wave' state transforms as the B$_{g}$ representation of the group, while the `d$_{xy}$-wave state again transforms as the trivial representation (where $x$ and $y$ are oriented at 45$^\circ$ to the crystal axes) \cite{BenGroupTh}.
\end{thebibliography}
\end{document}